\documentclass[10pt]{article}

\usepackage{arxiv}

\usepackage[utf8]{inputenc} 
\usepackage[T1]{fontenc}    
\usepackage[hidelinks]{hyperref}       
\usepackage{url}            
\usepackage{booktabs}       
\usepackage{amsfonts}       
\usepackage{nicefrac}       
\usepackage{microtype}      
\usepackage{xcolor}
\usepackage{graphicx}
\usepackage{subcaption}
\usepackage{multicol}
\usepackage[ruled,vlined]{algorithm2e}

\usepackage{tablefootnote}

\newenvironment{Figure}
  {\par\medskip\noindent\minipage{\linewidth}}
  {\endminipage\par\medskip}
  
\newenvironment{Table}
  {\par\medskip\noindent\minipage{\linewidth}}
  {\endminipage\par\medskip}

\title{BioKlustering: a web app for semi-supervised learning of maximally imbalanced genomic data}

\author{
Samuel Ozminkowski\thanks{Joint first authors with equal contribution} \\
Department of Statistics \\
University of Wisconsin-Madison
\And 
Yuke Wu$^*$ \\
Department of Computer Science \\
University of Wisconsin-Madison
\And 
Hailey Bruzzone \\
Department of Statistics \\
University of Wisconsin-Madison
\And
Liule Yang \\
Department of Computer Science \\
University of Wisconsin-Madison
\And
Zhiwen Xu \\
Department of Statistics \\
University of Wisconsin-Madison
\And
Luke Selberg \\
Department of Electrical and Computer Engineering \\
University of Wisconsin-Madison
\And
Chunrong Huang \\
Department of Computer Science \\
University of Wisconsin-Madison
\And
Helena Jaramillo Mesa\\
Department of Plant Pathology \\
University of Wisconsin-Madison
\And 
Claudia Sol\'is-Lemus\thanks{Corresponding author: solislemus@wisc.edu}\\
Wisconsin Institute for Discovery \\
Department of Plant Pathology \\
University of Wisconsin-Madison
}

\begin{document}
\maketitle




\vspace{0.5cm}

\begin{abstract}
Accurate phenotype prediction from genomic sequences is a highly coveted task in biological and medical research. While machine-learning holds the key to accurate prediction in a variety of fields, the complexity of biological data can render many methodologies inapplicable.
We introduce BioKlustering, a user-friendly open-source and publicly available web app for unsupervised and semi-supervised learning specialized in cases when sequence alignment and/or experimental phenotyping of all classes are not possible. 
Among its main advantages, BioKlustering 1) allows for maximally imbalanced settings of partially observed labels including cases when only one class is observed, which is currently prohibited in most semi-supervised methods, 2) takes unaligned sequences as input and thus, allows learning for widely diverse sequences (impossible to align) such as virus and bacteria, 3) is easy to use for anyone with little or no programming expertise, and 4) works well with small sample sizes.
%
%
BioKlustering (\url{https://bioklustering.wid.wisc.edu}) is a freely available web app implemented with Django, a Python-based framework, with all major browsers supported. The web app does not need any installation, and it is publicly available and open-source (\url{https://github.com/solislemuslab/bioklustering}).
%
%
%
\end{abstract}


\begin{multicols}{2}
\section{Introduction}

\noindent \textbf{Background.} The accurate prediction of biological features from genomic data is paramount for precision  medicine,  sustainable  agriculture  and  climate  change  research. 
Yet some characteristics of big biological data render most out-of-the-box machine-learning methodologies inaccurate or inapplicable. 
In particular, here we focus on 1) the high complexity (and sometimes impossibility) to align genomic sequences for certain organisms, and 2) the difficulty to obtain labeled samples to use in supervised learning.
First, while the growing interest of the biological community in machine-learning methods is undeniable, many existing machine-learning methods \cite{Agarwal2019-pr, Pei2018-ei, wnuk2018predicting, greenside2018discovering, zhang2016deepsplice} 
 need DNA or RNA alignments as input.
Fast evolving organisms like virus or bacteria complicate the alignment process with their heterogeneity and genomic diversity to the point that aligning the sequences becomes extremely challenging, if not impossible \cite{ren2018alignment, thompson2011comprehensive}. This is even more challenging when we want to include large number of sequences from distantly related groups.
Second, supervised learning models are the most accurate options for phenotype prediction. However, these methods rely on large samples of labeled data. In reality, many biological experiments produce partially observed labels given the money and time constraints to phenotype organisms or strains. One example that motivated our work is the case of mycovirus. Mycoviruses infect fungi and can be used as biocontrol of crop pests given that they can induce hypovirulence on plant fungal pathogens \cite{garcia2019mycoviruses}. Not all mycoviruses cause hypovirulence in the fungal host, and thus, a standard mycovirus dataset will be partially labeled with some labeled sequences that have been tested \textit{in planta} for their hypovirulence potential, and many more unlabeled sequences that have never been tested in laboratory, and thus, have unknown hypovirulence potential. 
Semi-supervised methods \cite{doostparast2018graph, bhardwaj2010genome} allow clustering based on partially observed labels, but existing methods require that we observe labels for all classes -- albeit in smaller frequencies than the unobserved labels. This condition is not met in the mycovirus dataset, for example, where negative controls (class of mycoviruses that do not cause host hypovirulence) are generally not available as observed labels only correspond to mycoviruses tested in the lab for their hypovirulence-induced potential. Thus, the use of most semi-supervised methods is prohibited for this maximally imbalanced dataset.

\noindent \textbf{Main contributions.}
We present BioKlustering (\url{https://bioklustering.wid.wisc.edu}) a user-friendly open-source web app to cluster unaligned genomic sequences based on maximally imbalanced partially observed labels. 
Unlike most semi-supervised learning methods, our web app does not require all classes to be observed or has any requirements on class balance. Also, unlike most machine-learning methods, our web app does not require large sample sizes. While designed for a semi-supervised setup, our web app can also be used when no labels are measured (unsupervised case).
One of the main purposes of our web app is to allow easy and fast clustering of sequences that could inform future biological experiments. Indeed, we see our web app as a hypothesis-generating tool that will allow biological users, for example, to identify clusters of mycovirus sequences that could be either negative controls (from the unobserved class of mycovirus that do not serve as biocontrol and that have never been tested \textit{in planta}) or new mycovirus sequences that could serve as biocontrol for further experimentation.
We highlight that even when our work is not inferential and mainly tailored for visualization purposes, we believe that it fills a important gap for biological scientists who need to identify similar sequences for experiments when sequence alignment is prohibited and when \textit{in vitro} or \textit{in planta} phenotyping of all labels is not possible. 


\section{Methods}

\noindent \textbf{Input data.}
Sequences are input as FASTA files and are internally converted to kmer counts. For example, if a sequence is ``ACTGG", then its 3-mers are [``ACT", ``CTG", ``TGG"]. The user can select the length of the kmer and sequences can be aligned or unaligned.
Labels are input as an optional separate csv file. The matching is done by assuming that the rows in both files are in the same order. The labels file should assign a value of $-1$ to the sequences of unknown label.


\noindent \textbf{Models.} We implement three unsupervised clustering methods which are extended to the case of partially observed labels by internally optimizing the kmer length and other parameters until we reach maximum consensus with the observed labels.

\noindent \textit{K-means clustering.} The algorithm \cite{macqueen1967some} groups the data into $K$ clusters each formed around a centroid. Each data point is assigned to the nearest centroid, and these centroids are formed by minimizing the squared Euclidian distances within each cluster. The mean-shift algorithm is used to identity locations of high density within the kmer space of the data, and then the unsupervised k-means model is run with these locations as the initial centroid coordinates.
Parameters of this model include the minimum and maximum kmer length, the random seed, and the number of clusters. Since the mean-shift algorithm does not allow an input number of clusters, when the number of clusters predicted by mean-shift exceeds the number of clusters requested, we use Algorithm 1 (Appendix) to assign labels. This algorithm combines the smallest extra clusters to the same label (for example, if $N$ clusters are requested and $M > N$ clusters are produced by the algorithm, the $M-N+1$ smallest clusters are all given label $N$).
For the semi-supervised case, we use the Algorithm 2 (Appendix) to map estimated output labels to the original input labels.


\noindent \textit{Gaussian Mixture Model.} The algorithm \cite{Buitinck2013-dn} fits a probabilistic model that estimates the multiple Gaussian distributions that best describe the clusters in the data. Starting with a random initialization model parameters, the Expectation-Maximization algorithm runs iteratively until convergence. 
Parameters of this model include the minimum and maximum kmer length, the random seed, the number of classes, and the covariance type (which determines the shape of the clusters). 
Again, for the semi-supervised case, we use the Algorithm 2 (Appendix) to map estimated output labels to the original input labels.

\noindent \textit{Spectral clustering.}
This algorithm \cite{pentney2005spectral} exploits the potential of eigenvalues of the matrix derived from the input data. In spectral clustering, the input data will be treated as a graph, and each sequence will represent a vertex in the graph. Then, the vertices in the graph will be partitioned based on their similarities. 
Parameters of this model are the minimum and maximum kmer length, the random seed, the number of classes, and the manner in which we assign labels when the input is projected to a lower dimension space (denoted label assignment option).
We again use Algorithm 2 (Appendix) to map estimated output labels to the original input labels in the semi-supervised setting.

For all three algorithms, the predicted labels and embedded kmer data can be used to visualize the clusters using principle component analysis (PCA) or t-distributed stochastic neighbor embedding (t-SNE).

\begin{figure*}
    \centering
    \includegraphics[scale=0.17]{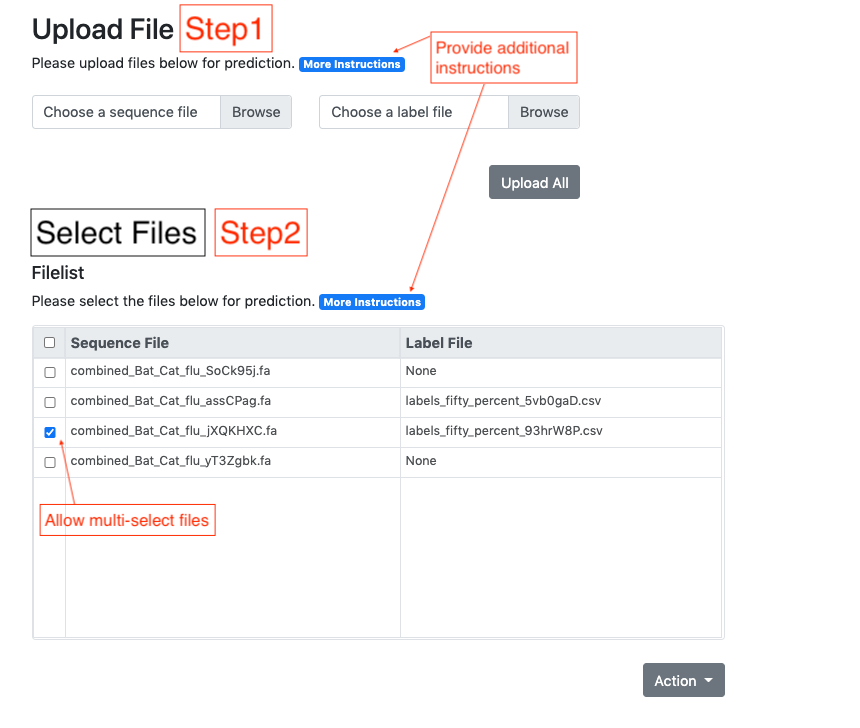}
    \includegraphics[scale=0.15]{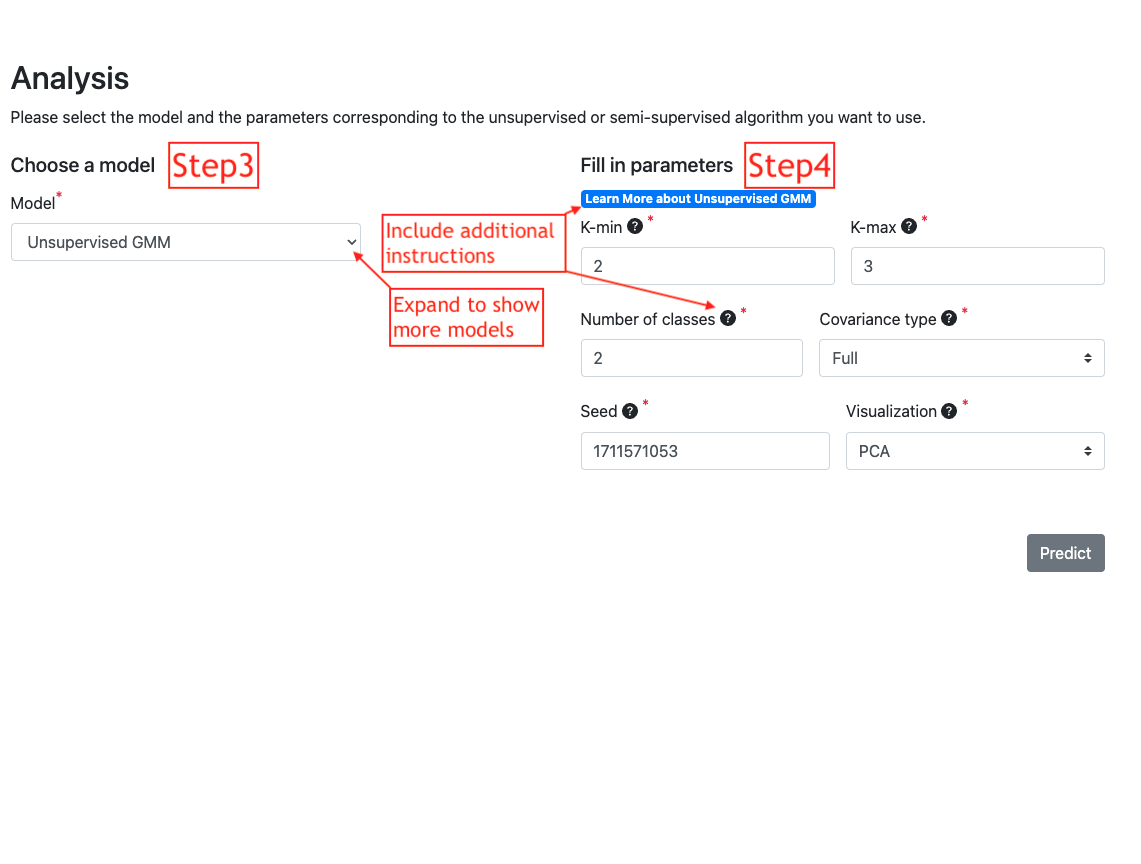}
    \includegraphics[scale=0.18]{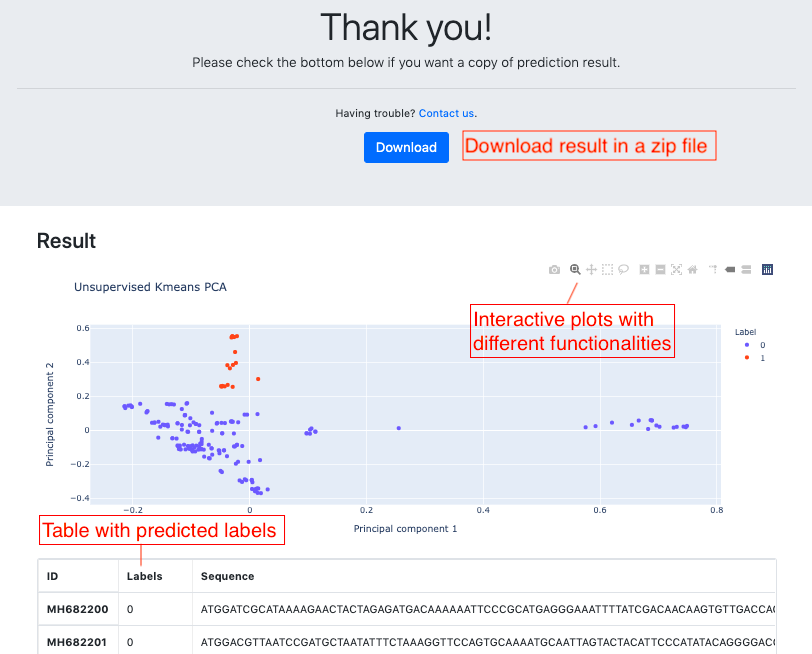}
    \captionof{figure}{Left: The Upload File section contains two file upload boxes, one for sequence file and one for label file. The uploaded files will be saved to the file list. Center: The Predict section contains a drop-down menu that displays the three clustering algorithms at the left-hand side and a parameter form at the right-hand side. Right: The Result page contains an interactive plot built with Plotly Dashboard and a table with predicted labels. }
    \label{webapp}
\end{figure*}

\paragraph{Web app interface.}
BioKlustering is an open-source web application developed with Django, a Python-based framework (Figure \ref{webapp}). Users will be able to predict clusters in genomic data by 1) uploading FASTA files with (aligned/unaligned) genome sequences, 2) selecting a clustering algorithm, and 3) choosing in its corresponding parameters. 

BioKlustering provides semi-supervised and unsupervised options depending on the presence of known labels. For each algorithm, the web app presents a description to aid users in the choice of parameters. The selected algorithm will generate prediction results which include an interactive plot (built with Plotly Dashboard \cite{plotly}) and a table with predicted labels. Users can download a static version of the plot, the table, and the parameter information in a zip file. 

\section{Validation} 

\subsection{Comparison with BINGO-3C on \textit{Influenza} data}

We downloaded \textit{Influenza} virus nucleotide sequences from the NCBI database \cite{virus-database}: 64 sequences from bat hosts with lengths between 844 and 2339 bp, and 114 sequences from cat hosts with lengths between 538 and 2341 bp. 

For this data, we have the labels for all sequences (whether they are from bat or cat hosts), but we run our learning algorithms under four settings: 1) unsupervised (ignoring all labels), 2) semi-supervised with 50\% of observed labels randomly selected, 3) semi-supervised with 10\% of observed labels randomly selected, and 4) semi-supervised with 10\% of observed labels from one class only. Out of the 50\% observed labels, 36\% of them correspond to the class of bats and 64\% to the class of cats. Out of the 10\% observed labels, 41\% of them correspond to the class of bats and 59\% to the class of cats when both classes are observed, and 100\% correspond to the class of bats when only one class is observed. 

We compared the performance of BioKlustering with BINGO-3C \cite{bingo3c}, a novel clustering method for genomic sequences that relies on a smarter embedding compared to the naive kmer count implemented in BioKlustering. The limitation of BINGO-3C, however, is that it does not provide a semi-supervised version (as BioKlustering does), so it would discard information in the case of some known labels. 
We modified BINGO-3C to recover the similary matrix and use it as input in the same three clustering algorithm as BioKlustering: k-means, GMM and spectral.

Table \ref{tab:pred} shows the prediction accuracy of the three methods: k-means, GMM and spectral clustering. See the Appendix for more details about the specific parameter choices for each model. Given the more sophisticated embedding in BINGO-3C compared to BioKlustering, it is expected that it will perform better in the unsupervised settings (first and fourth rows). However, when there is information on some labels (semi-supervised settings), BioKlustering outperforms BINGO-3C in most cases, except for the scenario where labels of only one class are observed. This scenario, however, still has accurate results.
Note that for GMM and spectral clustering, accuracy in the 10\% labeled with only 0s cases is worse than in the unsupervised case. This is also true for the regular 10\% labeled case for spectral clustering. This may seem counter-intuitive -- why should the model perform worse when it has more information? The problem here is analogous to overfitting: when only a few labels are provided, they carry a lot of weight, and in some cases the model performs better overall when a subset of these provided labels are removed. See the Appendix for a toy example illustrating this scenario.

\setlength{\tabcolsep}{3pt}
\begin{Table}
    \centering
    \begin{tabular}{l|ccc}
         & k-means & GMM & Spectral \\
         \hline
        Unsupervised & 71.3 & 75.3 & 75.8 \\
        Semi-supervised (50\%) & 82.6 & 93.8 & 87.1 \\
        Semi-supervised (10\%) & 73.6 & 90.4 & 75.3 \\
        Semi-supervised (10\% 0's) & 71.3 & 51.1 & 70.2 \\
        BINGO-3C (unsupervised) & 81.5 & 79.8 & 77.0        
    \end{tabular}
    \captionof{table}{Prediction accuracy (\%) of the three learning algorithms in BioKlustering and BINGO-3C on the \textit{Influenza} unaligned genome data for two classes (bats and cats) with respect to all the true labels.}
    \label{tab:pred}
\end{Table}

Table \ref{tab:pred-sm2} shows the prediction accuracy of the three clustering methods comparing only the observed labels. The observed labels are not modified by the prediction algorithm. That is, if a sample has an observed label of 0, it will always keep that label. In this table, however, we show the prediction accuracy if we pretend we did not observe those labels. This accuracy reflects how well the algorithm aligns with the known information (observed labels), and we cannot compare to BINGO-3C in this setting because BINGO-3C discards any known labels and performs unsupervised learning.

\setlength{\tabcolsep}{3pt}
\begin{Table}
    \centering
    \begin{tabular}{l|ccc}
         & k-means & GMM & Spectral \\
         \hline
        Semi-supervised (50\%) & 77.5 & 100 & 74.2 \\
        Semi-supervised (10\%) & 76.5 & 100 & 76.5 \\
        Semi-supervised (10\% 0's) & 100 & 100 & 100
    \end{tabular}
    \captionof{table}{Prediction accuracy (\%) of the three learning algorithms in BioKlustering on the \textit{Influenza} unaligned genome data for two classes (bats and cats) with respect to the observed labels.}
    \label{tab:pred-sm2}
\end{Table}






\subsection{Performance of BioKlustering on Mycovirus data}

Whole-genome sequences of 366 viral strains of mycoviruses were downloaded from GenBank. The genomic sequences belong to the \textit{Hypoviridae} family which contains most Sclerotinia mycoviruses and it is the family with the largest number of viruses associated with hypovirulance in fungi.
Out of these sequences, 7 had already been tested \textit{in planta} for their potential to induce hypovirulence in \textit{Sclerotinia} \cite{garcia2019mycoviruses} (label 1), a common crop fungal pathogen group \cite{reich2023predicting}, and 9 that do not have biocontrol potential (label 0). Those 16 strains are the ones that have complete polyprotein sequences available.
The remaining of the sequences (350) of mycoviruses have not been tested as potential biocontrol, and thus, the dataset has $16/366 \approx 4\%$ observed labels.



Table \ref{tab:nucleotide_results} shows the predicted labels of positive (or negative) biocontrol from the three algorithms in BioKlustering. The predicted labels are plotted on the tips of a phylogenetic tree estimated with IQ-Tree \cite{nguyen2015iq} (Figure \ref{heatmap-tree}) where we can perceive clustering of biocontrol (red) vs non-biocontrol (blue) in closely related strains. The fact that colors segregate on clades provides validation that the algorithm is accurately clustering similar sequences.

\setlength{\tabcolsep}{3pt}
\begin{Table}
    \centering
    \begin{tabular}{c|c|c}
         & Positive & Negative \\\hline
        GMM & 201 & 165 \\
        KMeans & 25 & 341 \\
        Spectral & 207 & 159 \\
        All & 19 & 144
    \end{tabular}
    \captionof{table}{Predicted labels of biocontrol (positives) and not biocontrol (negatives) strains under the three algorithms in BioKlustering. The last row denoted ``All" represents the number of strains predicted to be positive (or negative) by all three algorithms.}
    \label{tab:nucleotide_results}
\end{Table}




\begin{Figure}
    \centering  
    \includegraphics[scale=0.7]{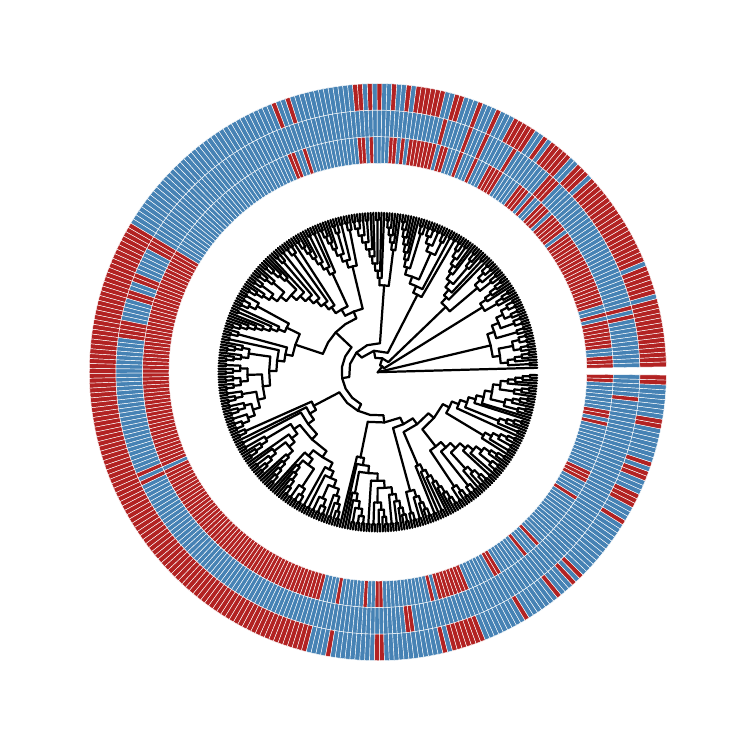}
    \captionof{figure}{Phylogenetic tree on the 366 mycovirus strains with predicted biocontrol labels (positive in red and negative in blue) at the tips for the three algorithms in BioKlustering (spectral in the outer circle, k-means in the center circle and GMM in the inner circle).}
    \label{heatmap-tree}
\end{Figure}

\noindent \textbf{Open-source code.} All of our code is open source in the following GitHub repository \url{https://github.com/solislemuslab/bioklustering}.

\noindent \textbf{Acknowledgements.} This work was supported by the Department of Energy [DE-SC0021016 to CSL]. 
We thank Ben Huebner from the WID IT team who helped with the public deployment of the web app. 
Finally, we acknowledge the work in \cite{Hotaling2020} which helped us improve the scientific writing of this manuscript.

\footnotesize{
\bibliographystyle{unsrt}  
\bibliography{references}  

\begin{thebibliography}{10}

\bibitem{Agarwal2019-pr}
Vishal Agarwal, N~Jayanth Kumar~Reddy, and Ashish Anand.
\newblock Unsupervised representation learning of {DNA} sequences.
\newblock {\em arXiv}, (1906.03087), June 2019.

\bibitem{Pei2018-ei}
Wenjie Pei and David M~J Tax.
\newblock Unsupervised learning of sequence representations by autoencoders.
\newblock April 2018.

\bibitem{wnuk2018predicting}
Kamil Wnuk, Jeremi Sudol, Kevin~B Givechian, Patrick Soon-Shiong, Shahrooz
  Rabizadeh, Christopher Szeto, and Charles Vaske.
\newblock Predicting dna accessibility in the pan-cancer tumor genome using
  rna-seq, wgs, and deep learning.
\newblock {\em bioRxiv}, page 229385, 2018.

\bibitem{greenside2018discovering}
Peyton Greenside, Tyler Shimko, Polly Fordyce, and Anshul Kundaje.
\newblock Discovering epistatic feature interactions from neural network models
  of regulatory dna sequences.
\newblock {\em Bioinformatics}, 34(17):i629--i637, 2018.

\bibitem{zhang2016deepsplice}
Yi~Zhang, Xinan Liu, James~N MacLeod, and Jinze Liu.
\newblock Deepsplice: Deep classification of novel splice junctions revealed by
  rna-seq.
\newblock In {\em 2016 IEEE international conference on bioinformatics and
  biomedicine (BIBM)}, pages 330--333. IEEE, 2016.

\bibitem{ren2018alignment}
Jie Ren, Xin Bai, Yang~Young Lu, Kujin Tang, Ying Wang, Gesine Reinert, and
  Fengzhu Sun.
\newblock Alignment-free sequence analysis and applications.
\newblock {\em Annual Review of Biomedical Data Science}, 1:93--114, 2018.

\bibitem{thompson2011comprehensive}
Julie~D Thompson, Benjamin Linard, Odile Lecompte, and Olivier Poch.
\newblock A comprehensive benchmark study of multiple sequence alignment
  methods: current challenges and future perspectives.
\newblock {\em PloS one}, 6(3):e18093, 2011.

\bibitem{garcia2019mycoviruses}
MD~Garc{\'\i}a-Pedrajas, MC~Ca{\~n}izares, Jorge~Luis Sarmiento-Villamil,
  Andr{\'e}s~Gustavo Jacquat, and Jos{\'e}~Sebasti{\'a}n Dambolena.
\newblock Mycoviruses in biological control: From basic research to field
  implementation.
\newblock {\em Phytopathology}, 109(11):1828--1839, 2019.

\bibitem{doostparast2018graph}
Abolfazl Doostparast~Torshizi and Linda~R Petzold.
\newblock Graph-based semi-supervised learning with genomic data integration
  using condition-responsive genes applied to phenotype classification.
\newblock {\em Journal of the American Medical Informatics Association},
  25(1):99--108, 2018.

\bibitem{bhardwaj2010genome}
Nitin Bhardwaj, Mark Gerstein, and Hui Lu.
\newblock Genome-wide sequence-based prediction of peripheral proteins using a
  novel semi-supervised learning technique.
\newblock {\em BMC bioinformatics}, 11(1):1--8, 2010.

\bibitem{macqueen1967some}
James MacQueen et~al.
\newblock Some methods for classification and analysis of multivariate
  observations.
\newblock In {\em Proceedings of the fifth Berkeley symposium on mathematical
  statistics and probability}, volume~1, pages 281--297. Oakland, CA, USA,
  1967.

\bibitem{Buitinck2013-dn}
Lars Buitinck, Gilles Louppe, Mathieu Blondel, Fabian Pedregosa, Andreas
  Mueller, Olivier Grisel, Vlad Niculae, Peter Prettenhofer, Alexandre
  Gramfort, Jaques Grobler, Robert Layton, Jake Vanderplas, Arnaud Joly, Brian
  Holt, and Ga{\"e}l Varoquaux.
\newblock {API} design for machine learning software: experiences from the
  scikit-learn project.
\newblock {\em arXiv}, (1309.0238), 2013.

\bibitem{pentney2005spectral}
William Pentney and Marina Meila.
\newblock Spectral clustering of biological sequence data.
\newblock In {\em AAAI}, volume~5, pages 845--850, 2005.

\bibitem{plotly}
Mark Gibbs.
\newblock django-plotly-dash 1.5.0
  \url{https://pypi.org/project/django-plotly-dash/}.
\newblock Date accessed: December 2020.

\bibitem{virus-database}
https://www.ncbi.nlm.nih.gov/genomes/FLU/Database/nph select.cgi.
\newblock Influenza virus database.
\newblock 2020 (accessed July 24, 2020).

\bibitem{bingo3c}
Bingo-3c.
\newblock \url{https://github.com/mouneem/BiNGO-3C/tree/main}.
\newblock Accessed: 2023-09-16.

\bibitem{reich2023predicting}
Jonathan Reich and Syama Chatterton.
\newblock Predicting field diseases caused by sclerotinia sclerotiorum: A
  review.
\newblock {\em Plant Pathology}, 72(1):3--18, 2023.

\bibitem{nguyen2015iq}
Lam-Tung Nguyen, Heiko~A Schmidt, Arndt Von~Haeseler, and Bui~Quang Minh.
\newblock Iq-tree: a fast and effective stochastic algorithm for estimating
  maximum-likelihood phylogenies.
\newblock {\em Molecular biology and evolution}, 32(1):268--274, 2015.

\bibitem{Hotaling2020}
Scott Hotaling.
\newblock Simple rules for concise scientific writing.
\newblock {\em Limnology and Oceanography Letters}, 5(6):379--383, 2020.

\end{thebibliography}
}

\end{multicols}

\pagebreak

\appendix

\section{Algorithms}

When more clusters are requested than the k-means algorithm generates, we must combine some to reduce the number of clusters. To do this, we combine the smallest clusters (based on number of data points) into a single cluster. For example, if $N$ clusters are requested and $M>N$ clusters are generated by the k-means algorithm, the $M-N+1$ smallest clusters are combined into cluster $N$. See Algorithm \ref{algo:assignClustersKM} for more details.

\begin{algorithm}[H]
\caption{K-Means Cluster Label Assignment}
\label{algo:assignClustersKM}

\KwResult{Labeled data}
\If{number of clusters predicted $\geq$ number of clusters requested}{
    N $\gets$ number of clusters requested\;
    Sort clusters by size\;
    \For{ i $\in \mathbb{Z}$, i $\leq$ number of clusters predicted }
    {
        \eIf{ $i \leq$  number of clusters requested}
        {
            assign label $i$ to points in sorted cluster $i$ \;
        }{
            assign label N to points in sorted cluster $i$\;
        }
    }
}
\end{algorithm}

With semi-supervised labeling, the labels produced by the clustering algorithm may differ from the inputted labels. In order to correct this, we map labels produced by the algorithm (predicted labels) to labels given by the user (given labels). This is done greedily by matching the largest (by number of data points) unassigned given label to the predicted label with the most data points (of those with pre-assigned labels) matching said given label. See Algorithm \ref{algo:assignClusters} for more details.

\begin{algorithm}[H]
\caption{Cluster Label Assignment}
\label{algo:assignClusters}

\KwResult{Labeled data}
MAX\_VALUE $\gets$ $\max$(labels input) + 1
\\GIVEN\_LABELS\_CT $\gets$ dictionary: \{given label $\to$ \#\{given labels\}\}, sorted by value
\\PRED\_LABELS\_CT $\gets$ dictionary: \{predicted label $\to$ \#\{predicted labels\}\}, sorted by value
\\UNSELECTED\_PRED $\gets$ PRED\_LABELS\_CT
\\map\_predict\_to\_actual $\gets$ empty dictionary
\\\For{given\_label $\in$ GIVEN\_LABELS\_CT }
{
    pred\_results $\gets$ predicted labels whose true label matches given\_label
    \\ unique\_predicted $\gets$ unique(pred\_results) $\cap$ UNSELECTED\_PRED
    \\ \If{ unique\_predicted $= \emptyset$ }
    {
        continue
    }
    predicted\_labels\_ct $\gets$ dictionary: \{unique\_predicted $\to$ \#\{unique\_predicted = pred\_results\}\}
    \\ map\_predict\_to\_actual[$\max$ (unique\_predicted)] $\gets$ given\_label
    \\ remove $\max$ (unique\_predicted) from UNSELECTED\_PRED
}

\For{unique\_predicted\_label\_unselected $\in$ UNSELECTED\_PRED}
{
    map\_predict\_to\_actual[unique\_predicted\_label] $\gets$ MAX\_VALUE
    \\ MAX\_VALUE $\gets$ MAX\_VALUE + 1
}
output $\gets$ []
\\\For{i $\in$ \{0,...,\#{predictions}\}}
{
    \eIf{\textrm{no label was input for index i}}
    {
        output[i] $\gets$ map\_predict\_to\_actual[predicted label i]
    }
    {
        output[i] $\gets$ input[i]
    }
}

\end{algorithm}

\section{Validation on \textit{Influenza} data}

We test our three semi-supervised algorithms with \textit{Influenza} data from two classes: bats and cats \cite{virus-database}. We run four settings: totally unsupervised, semi-supervised with 50\% of observed labels (both classes observed), semi-supervised with 10\% observed labels (both classes observed) and semi-supervised with 10\% observed labels (only one class observed). 

For the k-means method, the chosen parameters that maximize the agreement with the observed labels are $k_{min}=5,k_{max}=7$ in all four cases.

For the GMM method, the chosen parameters that maximize the agreement with the observed labels are 
\begin{itemize}
    \item $k_{min}=2,k_{max}=6$ with covariance type as ``full" for the unsupervised case,
    \item $k_{min}=4,k_{max}=5$ with covariance type as ``diagonal" for the semi-supervised case with 10\% labels observed (both classes),
    \item $k_{min}=2,k_{max}=5$ with covariance type as ``full" for semi-supervised case with 10\% observed labels (only 0's),
    \item $k_{min}=3,k_{max}=4$ with covariance type as "tied" for the semi-supervised case with 50\% labels observed.
\end{itemize}

For the spectral clustering, the chosen parameters that maximize the agreement with the observed labels are 
\begin{itemize}
    \item $k_{min}=2,k_{max}=3$ with k-means label assign option for the unsupervised case,
    \item $k_{min}=k_{max}=3$ with discretize label assign option for the semi-supervised case with 10\% labels observed (both classes),
    \item $k_{min}= k_{max}=2$ and label assign option as discretize for when only 0's are observed,
    \item there were two setups with the same accuracy for the semi-supervised case with 50\% observed labels: $k_{min}=k_{max}=2$ with k-means label assign option and $k_{min}=3,k_{max}=5$ with discretize label assign option.
\end{itemize}

In all cases, the true number of classes is selected (2) and we do not test the performance when the number of clusters is not known in advance.

\begin{figure}[h]
    \centering
    \includegraphics[scale=0.25]{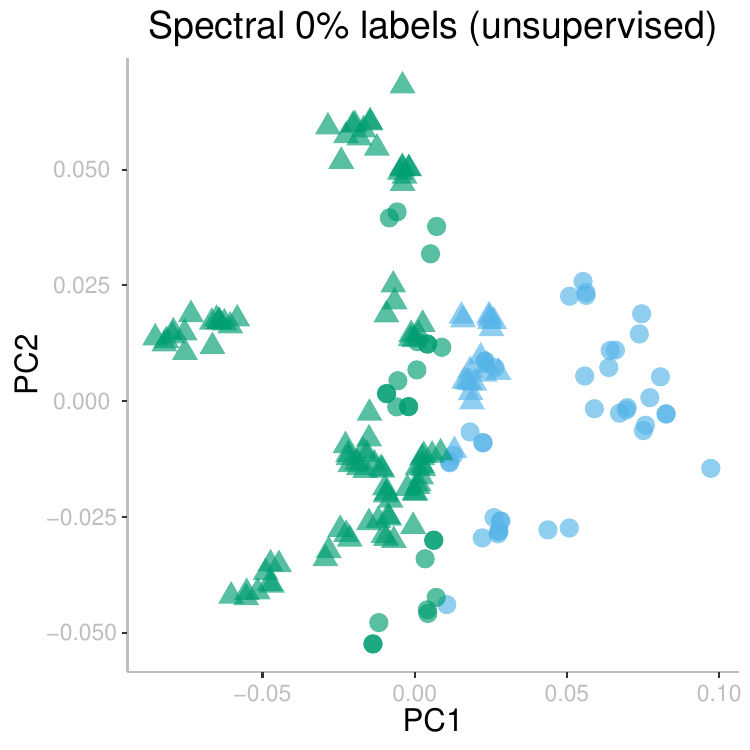}
    \includegraphics[scale=0.25]{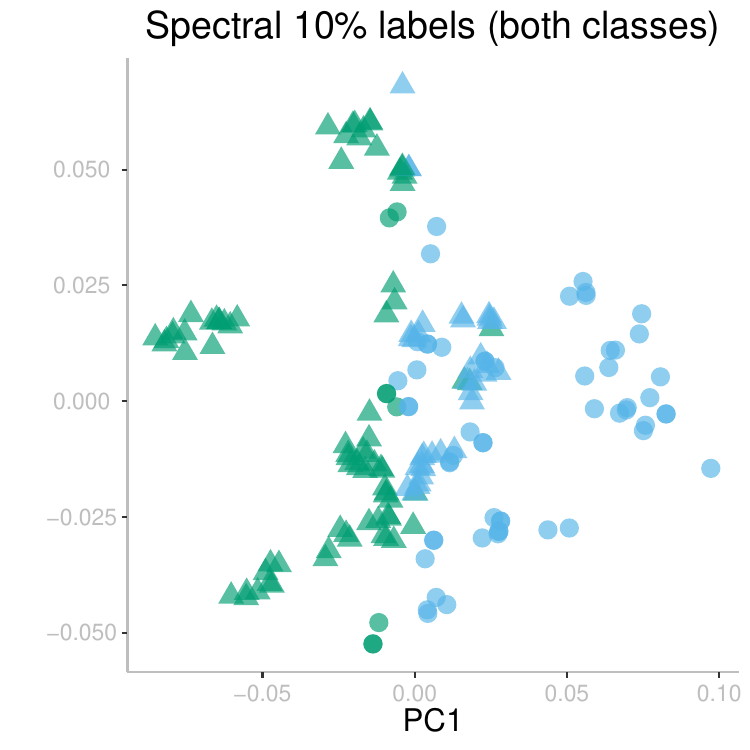}
    \includegraphics[scale=0.25]{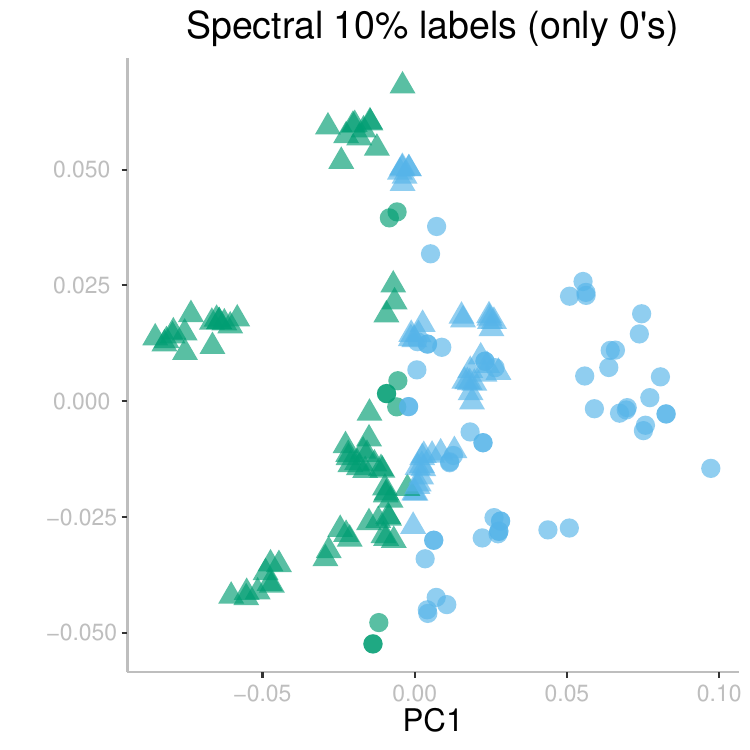}
    \includegraphics[scale=0.25]{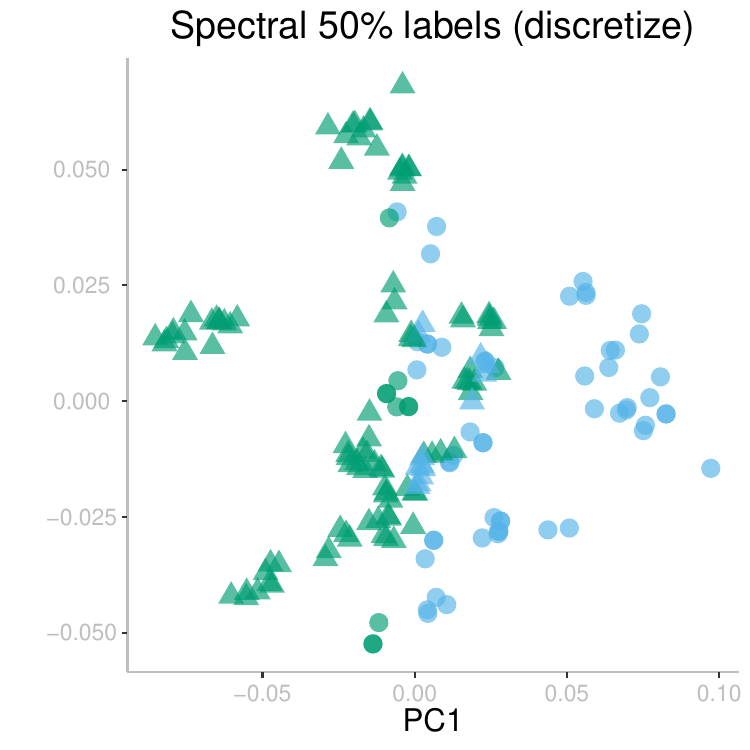}
    \includegraphics[scale=0.25]{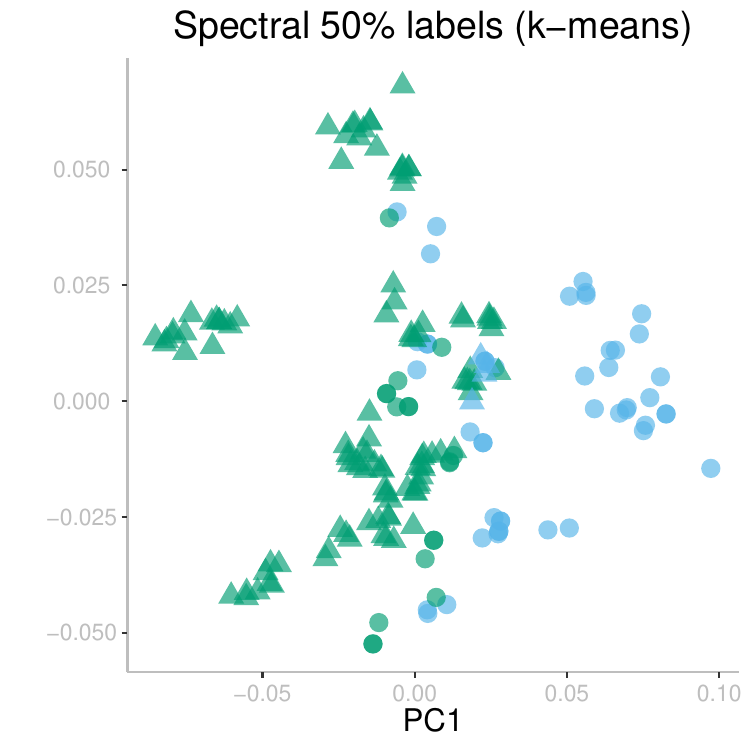} \\[0.5cm]
    \includegraphics[scale=0.25]{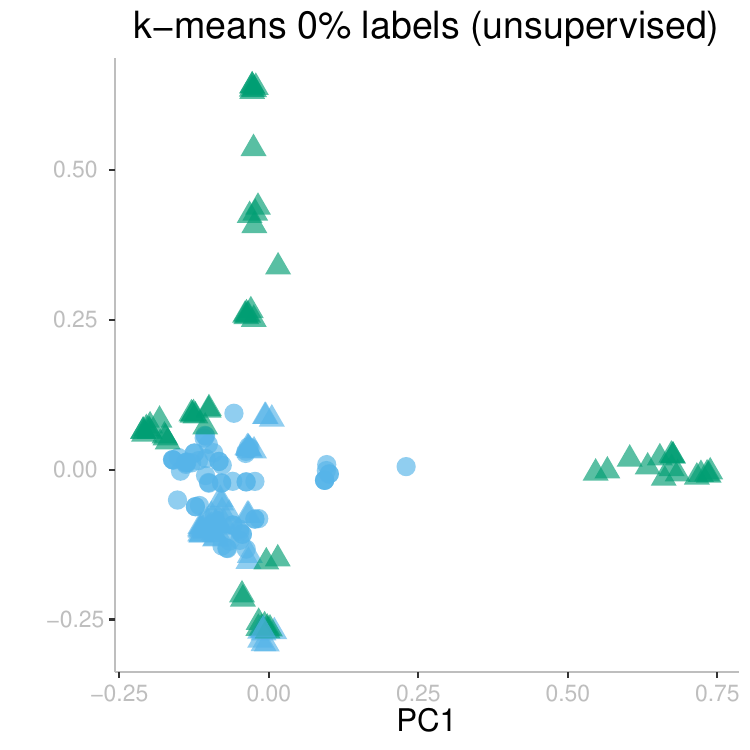}
    \includegraphics[scale=0.25]{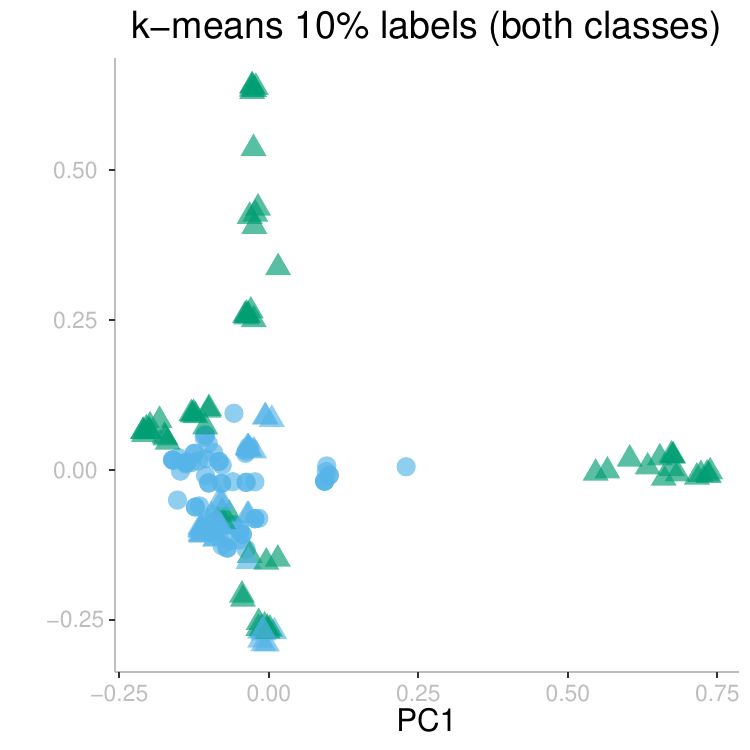}
    \includegraphics[scale=0.25]{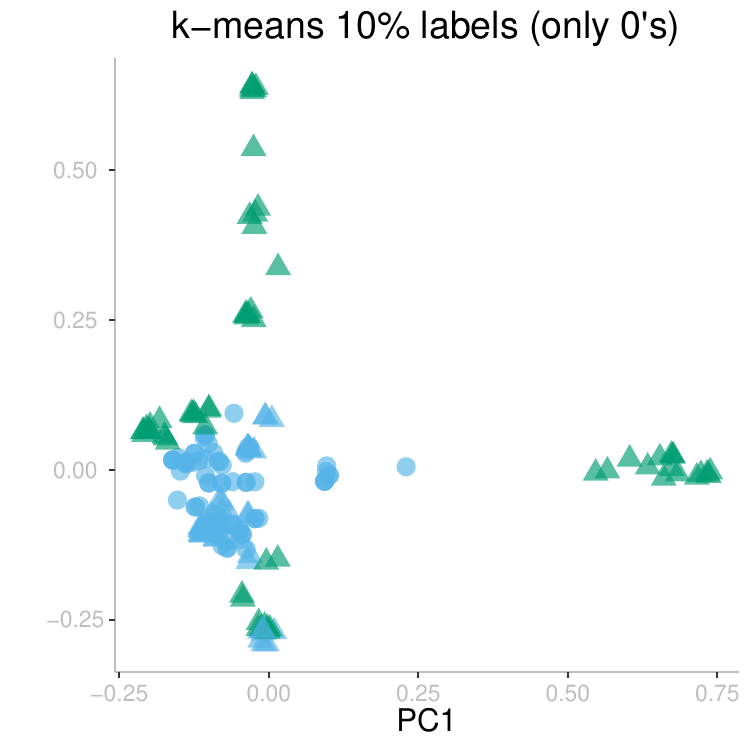}
    \includegraphics[scale=0.25]{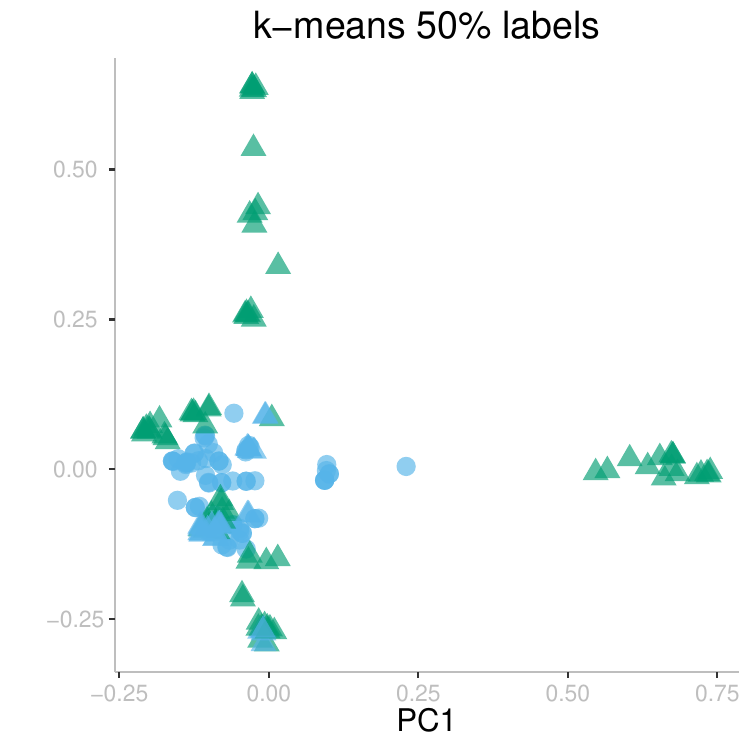} \\[0.5cm]
    \includegraphics[scale=0.25]{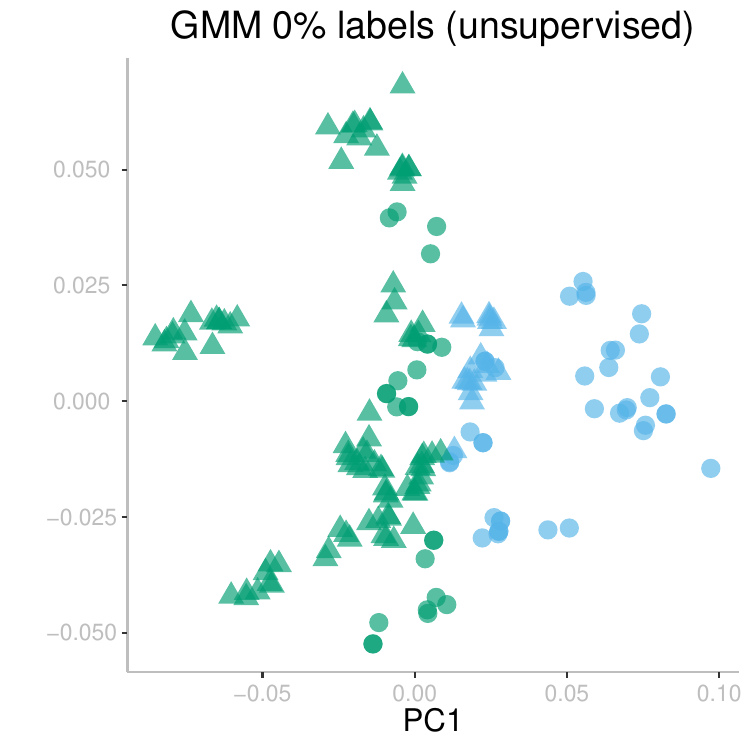}
    \includegraphics[scale=0.25]{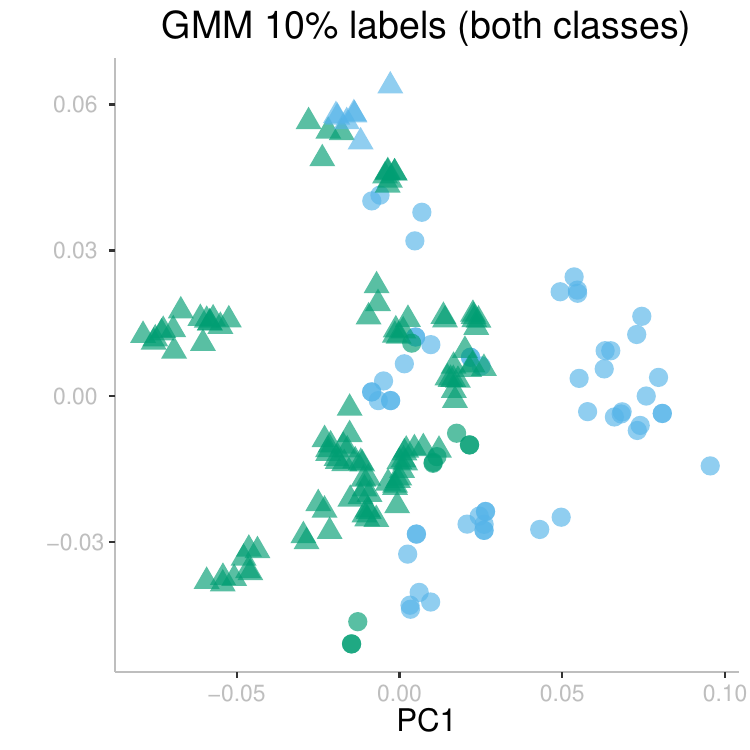}
    \includegraphics[scale=0.25]{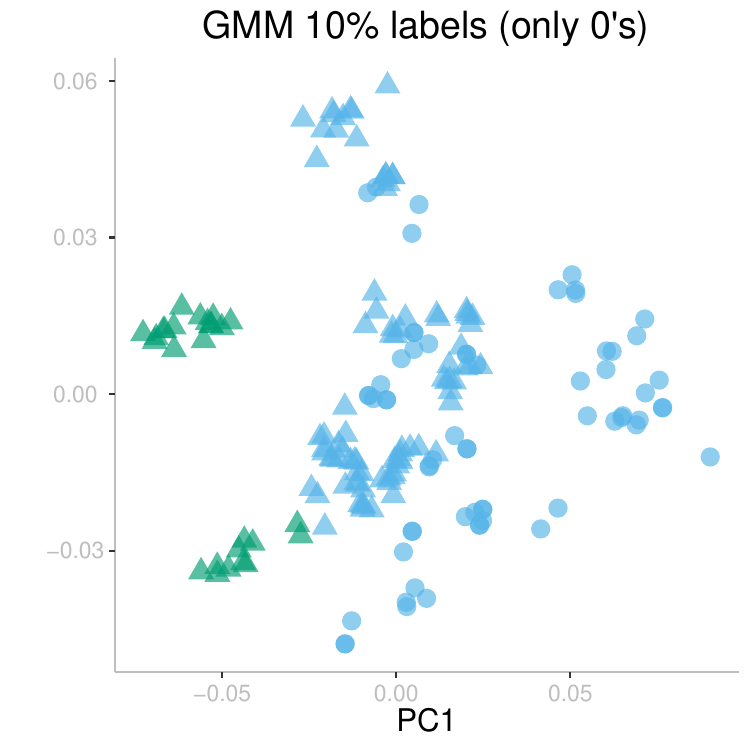}
    \includegraphics[scale=0.25]{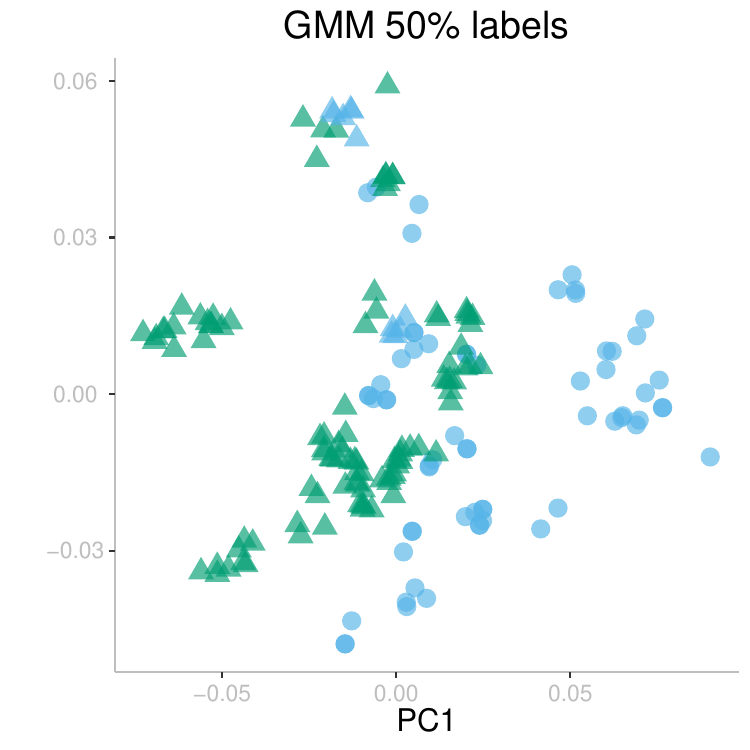}
    \caption{Clustering visualization of bats and cats \textit{Influenza} data for the three methods in BioKlustering. \textbf{Top:} spectral clustering. \textbf{Middle:} k-means clustering. \textbf{Bottom:} GMM clustering. There are three settings per method: unsupervised (no labels observed), semi-supervised with 10\% labels observed (both classes or only one class observed) and semi-supervised with 50\% labels observed. Color represents predicted labels: blue for bats (label 0) and green for cats (label 1). Point shapes represent true labels: circle for bats (label 0) and triangle for cats (label 1). Blue circles or green triangles correspond to correctly predicted labels, while green circles and blue triangles correspond to wrongly estimated labels.}
    \label{fig:all}
\end{figure}

Figure \ref{fig:all} shows the clustering visualization via PCA of the spectral clustering (top), the k-means clustering (middle), and the GMM clustering (bottom).

\section{Toy example on unsupervised vs semi-supervised performance}

Consider the toy example in Figure \ref{fig:toy_example}, where the blue and green points represent separate true groups and the only labeled points are the blue squares. In this case, a semi-supervised clustering algorithm would classify all points above the red line as blue and all points below as green, for an accuracy of approximately 13\%, while an unsupervised clustering would simply classify the two clusters as different clusters, and lacking any context as to which is which we would naively claim the accuracy as 87\%.

\begin{Figure}[ht]
    \centering
    \includegraphics[scale=0.8]{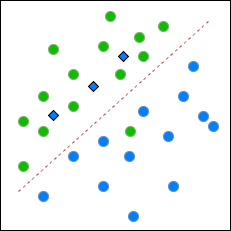}
    \captionof{figure}{Toy example showing why semi-supervised clustering with few labels can lead to worse results than unsupervised clustering. Colors represent the two true groups (green and blue) and shapes represent whether the label is known or unknown: squares are known labels and circles are unknown labels.}
    \label{fig:toy_example}
\end{Figure}

\end{document}